\documentclass[aps,pra,twocolumn,nofootinbib,superscriptaddress]{revtex4}

\usepackage{amsmath,amssymb,amsthm}
\usepackage{amsfonts}
\usepackage{graphicx}
\usepackage{comment}
\usepackage{color}

\newcommand{\blu}{\color{black}} %blue

\begin{document}

%\title{\blu Comment on ``Can Two-Way Direct Communication Protocols Be Considered Secure?"}
\title{On the security of two-way quantum key distribution}

\author{Jesni Shamsul Shaari}
\affiliation{Department of Physics, International Islamic University Malaysia (IIUM),
Jalan Sultan Ahmad Shah, Bandar Indera Mahkota, 25200 Kuantan, Pahang, Malaysia}
\affiliation{Institute of Mathematical Research (INSPEM), University Putra Malaysia, 43400 UPM Serdang, Selangor, Malaysia.}
\author{Stefano Mancini}
\affiliation{School of Science \& Technology, University of Camerino, I-62032 Camerino, Italy}
\affiliation{ INFN Sezione di Perugia, I-06123 Perugia, Italy}
\author{Stefano Pirandola}
\affiliation{Computer Science \& York Centre for Quantum Technologies,
University of York,
Deramore Lane, York, YO10 5GH, UK}
\author{Marco Lucamarini}
\affiliation{Toshiba Research Europe, 208 Cambridge Science Park, Cambridge CB4 0GZ, UK}

%\date{\today}

%\begin{abstract}
%Write an abstract here\end{abstract}

%\pacs{03.67.Dd, 03.65.Ta, 89.70.Cf}
%\keywords{Counterfactual QKD, CH violation}

\maketitle

%\section{Introduction}
\noindent In a recent article entitled ``Can Two-Way Direct Communication Protocols Be Considered Secure?"~\cite{pavi}, protocols for two-way Quantum Key Distribution (QKD)~\cite{tcs,PML+08}, {\blu in particular} the Ping-Pong {\blu protocol}~\cite{PP} and the LM05 {\blu protocol}~\cite{LM05}, were criticized under a specific {\blu well-known} attack {\blu called} ``Quantum-Man-in-the-Middle" (QMM)~\cite{nguyen,luc,tcs}. {\blu However, not only such an attack does not disprove the security of two-way QKD, but it also represents one of the weakest strategies available to an eavesdropper. }

{\blu Two-way QKD} involves a bidirectional quantum channel between {\blu the users} Alice and Bob. {\blu Bob sends qubits to Alice, who} either {\blu encodes} or {\blu measures them} in a randomly chosen basis. The former case corresponds to the encoding or message mode (MM), {\blu which is used to define the key;} the latter, to the control mode (CM), which {\blu is used to detect a potential} eavesdropper (Eve). Alice subsequently resubmits the qubits to Bob for his final measurements.  {\blu The quantum transmission is followed by a public discussion over an authenticated channel, where the users reconcile and distill the final keys}.

Ref.~\cite{pavi} {\blu overlooked the importance of the CM and} purported that two-way QKD {\blu becomes insecure} when attacks do not induce errors in the MM. The QMM attack {\blu  is used as a specific example. Here, Eve} swaps Bob's traveling qubit with her own, submits it to Alice and then upon learning Alice's {\blu encoding operation}, duplicates it onto Bob's qubit before returning it to him. Such an attack can only be detected in {\blu CM. This is enough for the author of Ref.~\cite{pavi} to conclude that: (i)} attacks which leave no errors in the MM {\blu do} not allow for security to be established by standard methods; {\blu (ii)} privacy amplification (PA) cannot be executed due to the absence of a `critical value' {\blu similar} to BB84's {\blu famous} $11\%$; {\blu (iii) the existing security proofs~\cite{hua} are flawed} as they do not consider this specific class of attacks.

We argue that {\blu the above} claims are erroneous {\blu and stem from a fundamental} misunderstanding of how {\blu two-way} QKD works.

{\blu This can be seen by considering the secure key rate of a QKD protocol. In the asymptotic limit of many signals, we can simply write it as $R=I_{AB}-I_{E}$, with $I_{AB}$ the mutual information between the users' keys and $I_{E}$ an upper bound on Eve's information on the key. In two-way QKD,} $I_E$ is obtained from the CM whereas $I_{AB}$ is estimated from the MM. Therefore, the CM makes it possible to properly execute the PA whereas the MM enables a proper execution of the EC.

The fact that the QMM attack introduces no error in the MM is inconsequential. {\blu In fact, while this implies that} $I_{AB}=1$, it also causes an error rate equal to 50\%, detected in the CM~\cite{nguyen,luc,tcs}. Hence, if Eve attacks a fraction $f$ of the qubits, the resulting secret key rate will be given by $R=1-f\geq 0$, with equality when $f=1$, i.e., when Eve attacks all the qubits.
Therefore, the protocol is always secure {\blu against such an attack, for any value} of $f$. The absence of errors in MM simply ensures the unity value for $I_{AB}$, to the legitimate parties' benefit, and has no bearing in determining the PA rate, which is a function of {\blu $I_{E}$} estimated from the error rate in CM. Obviously, differing values for errors in CM would translate into differing PA rates, hence differing $R$, effectively dismissing the notion of `critical value'. However, a similar argument applies even in the BB84 protocol, {\blu where the amount of PA executed depends on the actual error rate that has been detected, not on a predetermined critical value like} the $11\%$ quoted in Ref.~\cite{pavi}.

Finally, Ref.~\cite{pavi} also argued that the analysis in {\blu existing security proofs~\cite{hua}, which is based on} Eve's ancillae interacting with Bob's qubits via some unitary transformation, do not incorporate the {\blu QMM} attacking strategy. This is {\blu clearly untrue as QMM can be described as} a specific case of {\blu such proofs. It suffices to consider} Eve's ancilla as a qubit and the unitary transformation as the well known SWAP gate \cite{qc10}. An attack in the backward path is not made explicit as an extremely pessimistic stand is taken where Eve is allowed to extract all possible information from the entire Bob-Eve system without specifying the actual mechanism.
%It is also worth noting that in the same reference, there is no requirement set on $I_{AB}$ as an intrinsic part of the attack beyond what Alice and Bob must determine as part of post processing; thus a zero error rate in MM is completely acceptable without compromising the security analysis.

{\blu Two-way QKD security proofs are quite complex and include many different aspects. Simple attacks like those described in Ref.~\cite{pavi} have been important to move the first steps in this field, but real progress demands new avenues and better criticism. }\\
%\newline

\noindent {\bf Notes}: Let us remark we do not claim the security
of direct communication by the above, rather that of two-way QKD,
which is provably secure. It should be noted that despite its
title, the commented paper~\cite{pavi} was in fact addressing
two-way protocols within a QKD context.

\end{document}